\documentclass[aps,prb,amsmath,amssymb,twocolumn,superscriptaddress,floatfix]{revtex4-1}
\usepackage{helvet}
\usepackage{graphicx}
\usepackage{hyperref}
\hypersetup{pdftitle={Rh:EuFe2As2 Phase Diagram and ARPES},pdfauthor={Xiao Shaozhu},pdfsubject={Rh:EuFe2As2},pdfdisplaydoctitle}

\def\Eu{EuFe$_2$As$_2$}
\def\Ba{BaFe$_2$As$_2$}
\def\Sr{SrFe$_2$As$_2$}
\def\EuRhx{Eu(Fe$_{1-x}$Rh$_x$)$_2$As$_2$}
\def\EuRhopt{Eu(Fe\ensuremath{_{0.88}}Rh\ensuremath{_{0.12}})\ensuremath{_2}As\ensuremath{_2}}
\def\EuIropt{Eu(Fe\ensuremath{_{0.88}}Ir\ensuremath{_{0.12}})\ensuremath{_2}As\ensuremath{_2}}
\def\EuCaopt{(Eu\ensuremath{_{0.88}}Ca\ensuremath{_{0.12}})Fe\ensuremath{_2}As\ensuremath{_2}}
\def\Tc{\ensuremath{T_\text{c}}}
\def\TN{\ensuremath{T_\text{N}}}
\def\Hc2{\ensuremath{H_\text{c2}}}
\newcommand{\EuRh}[2]{Eu(Fe$_{#2}$Rh$_{#1}$)$_2$As$_2$} 
\begin{document}

\title{Electronic structure and {\itshape H}--{\itshape T} phase diagram of \EuRhx}

\author{Shaozhu Xiao}
\author{Darren C. Peets}
\email{dpeets@nimte.ac.cn}
\author{Wei Liu}
\author{Shiju Zhang}
\author{Ya Feng}
\affiliation{Ningbo Institute of Materials Technology and Engineering, Chinese Academy of Sciences, Ningbo, Zhejiang 315201, China}

\author{Wen-He Jiao}
\email{whjiao@zust.edu.cn}
\affiliation{Department of Physics, Zhejiang University of Science and Technology, Hangzhou, Zhejiang 310023, China}
\author{Guang-Han Cao}
\affiliation{Department of Physics, Zhejiang University, Hangzhou, Zhejiang 310027, China}
\affiliation{Collaborative Innovation Center of Advanced Microstructures, Nanjing, Jiangsu 210093, China}

\author{Eike F.\ Schwier}
\author{Kenya Shimada}
\affiliation{Hiroshima Synchrotron Radiation Center, Hiroshima University,
  Higashi-Hiroshima 739-0046, Japan}

\author{Cong Li}
\author{Xingjiang Zhou}
\affiliation{National Lab for Superconductivity, Beijing National Laboratory for Condensed Matter Physics, Institute of Physics, Chinese Academy of Sciences, Beijing 100190, China}

\author{Shaolong He}
\email{shaolonghe@nimte.ac.cn}
\affiliation{Ningbo Institute of Materials Technology and Engineering, Chinese Academy of Sciences, Ningbo, Zhejiang 315201, China}

\begin{abstract}

  The iron-based superconductors represent a promising platform for high-temperature superconductivity, but the interactions underpinning their pairing present a puzzle.  The \Eu\ family is unique among these materials for having magnetic order which onsets within the superconducting state, just below the superconducting transition.  Superconductivity and magnetic order are normally antagonistic and often vie for the same unpaired electrons, but in this family the magnetism arises from largely localized Eu moments and they coexist, with the competition between these evenly-matched opponents leading to reentrant superconducting behavior.  To help elucidate the physics in this family and the interactions between the magnetic order and superconductivity, we investigate the $H$--$T$ phase diagram near optimal Rh doping through specific heat, resistivity, and magnetization measurements, and study the electronic structure by angular-resolved photoemission spectroscopy. The competition between the Eu and FeAs layers may offer a route to directly accessing the electronic structure under effective magnetic fields via ARPES, which is ordinarily a strictly zero-field technique.

\end{abstract}

\maketitle

\section{Introduction}

In high-temperature superconductors, magnetism and superconductivity compete for control over the unpaired electrons, as exemplified by the competition of superconductivity with spin-density-wave and charge-density-wave order in the cuprates\cite{Ghiringhelli2012,Chang2012,Santi2013}. But the same high-temperature superconductivity may also owe its existence to fluctuations that arise as magnetic phases are suppressed --- in both the cuprate and iron-based superconductors, superconductivity appears where a phase with long-range spin order is suppressed, and that spin order otherwise persists to temperatures much higher than any superconducting critical temperature \Tc. More generally, the zero-field ground state of a superconductor with singlet pairing is perfectly diamagnetic, and magnetism within the material reduces the energy saved through pairing as the pair condensate is forced to adapt or compensate. Given the complex relationship between magnetism and high-temperature superconductivity, the interactions of magnetic order with these superconductors offers an important opportunity to shed light on the pairing in these systems, which remains among the greatest unsolved problems in condensed matter physics.

Doped \Eu\cite{Raffius1993,Ren2008,Jeevan2008} offers a unique opportunity to study this interaction.  It consists of high-temperature-superconducting FeAs layers, but the intermediate layer is a square lattice of magnetic Eu.  The FeAs and Eu layers both order magnetically in the undoped parent material\cite{Raffius1993}, but their magnetism is nearly decoupled\cite{Jeevan2008,Xiao2009}.  Superconductivity can be induced by pressure\cite{Matsubayashi2010} or by chemical substitution.  The latter can occur either on the Eu site, which frustrates the magnetic order while injecting charge carriers to the FeAs planes, or within the FeAs planes, which introduces significant disorder directly into the superconducting system but leaves the magnetic sublattice largely untouched.  Doped \Eu\ has a number of very unconventional properties, including a reentrant superconducting transition in the resistivity\cite{Jiao2017} and a reentrant spin-glass phase\cite{Baumgartner2017}, most likely as a result of the competition between superconductivity and magnetism.

Here we report the magnetic phase diagram and electronic structure of \EuRhx.  The superconductivity is not visible in the specific heat or ARPES data, likely as a result of the intrinsic competition between strong rare-earth magnetism in the Eu layer and high-temperature superconductivity in the FeAs layers.  This may offer a route to applying zero-field techniques such as ARPES to materials in an effective magnetic field.

\section{Experimental}\label{expt}

\begin{figure*}
  \includegraphics[width=\textwidth]{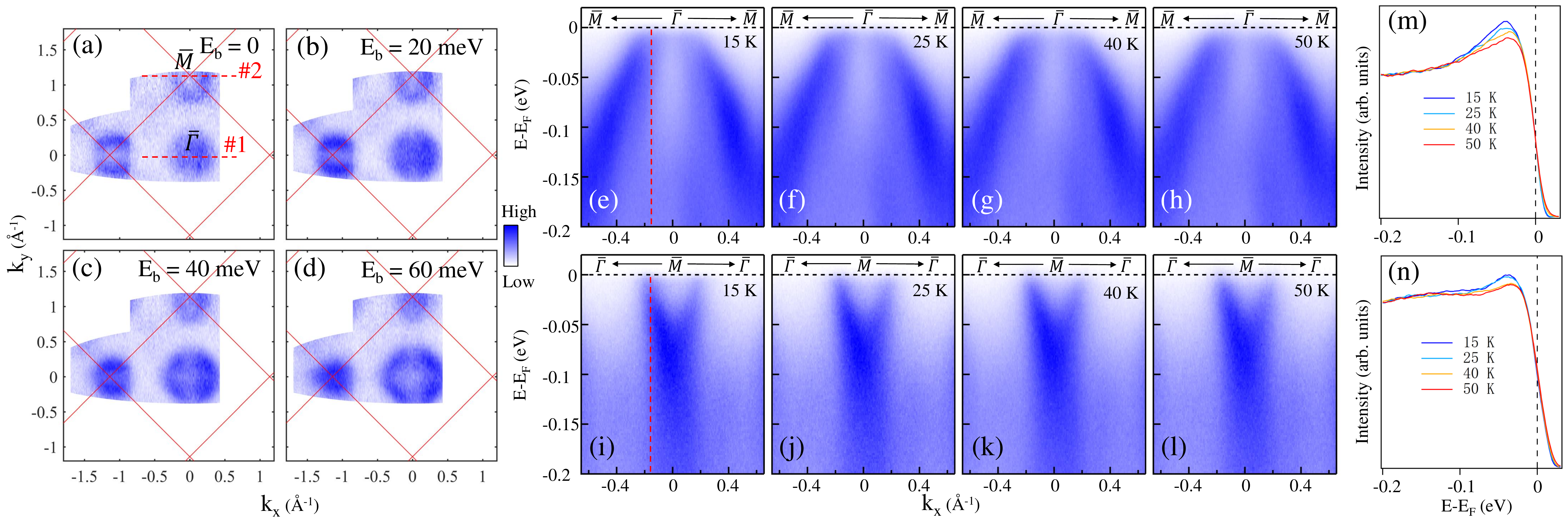}
  \caption{\label{fig:FSandBand}Electronic structure of \EuRhopt\ measured by ARPES with a photon energy $h\nu$ = 21.2\,eV. (a-d) Constant-energy contours at 15\,K at different binding energies as labeled. (e-l) Energy-momentum images taken at different temperatures along the (e-h) $\overline{M}$--$\overline{\Gamma}$--$\overline{M}$ direction as indicated by dashed line \#1 in (a), and (i-l) along the $\overline{\Gamma}$--$\overline{M}$--$\overline{\Gamma}$ direction as indicated by dashed line \#2 in (a). (m) Temperature-dependent EDCs extracted from images (e-h) at the Fermi momentum of the hole pocket, shown by the dashed line in (e). (n) Temperature-dependent EDCs extracted from images (i-l) at the Fermi momentum of the electron pocket, identified by the dashed line in (i).}
\end{figure*}

Single crystals of several dopings of \EuRh{x}{1-x}\ including undoped \Eu\ were grown from a (Fe,Rh)As self flux in Al$_2$O$_3$ crucibles as described elsewhere\cite{Jiao2013,Jiao2017}.  Measurements were performed chiefly on crystals with nominal composition \EuRh{0.12}{0.88}, which is near optimal doping.  Field-cooled (FC) and zero-field cooled (ZFC) magnetization measurements were performed in a Quantum Design MPMS3 vibrating sample SQUID magnetometer, with the crystal affixed to a quartz bar sample holder using GE Varnish and Teflon or Kapton tape.  Specific heat measurements were performed by the relaxation time technique in a Quantum Design Physical Properties Measurement System (PPMS), with the sample mounted using Apiezon N grease.  For in-plane fields, this crystal was attached to a small copper wedge to ensure both correct alignment and good thermal contact.  Several measurements were averaged for each data point, and in some temperature ranges it was necessary to discard the first measurement in each set because the sample's temperature had not fully equilibrated. Resistivity was measured using the standard four-wire technique in a Quantum Design PPMS, with a drive current $J$ of 5\,mA unless otherwise noted. The resistivity was measured for in-plane fields, which was accomplished by attaching the samples to slabs of sapphire which were then mounted upright on the sample puck.  Sample stoichiometry was verified by energy-dispersive x-ray spectroscopy (EDX). Samples with nominal Rh contents of 0.09, 0.12 and 0.16 were measured to be 0.068(3), 0.095(6) and 0.101(8) and have resistive \Tc s of 7.0, 18.6, and 18.8\,K, respectively. Since zero resistivity was not reached, superconducting transitions were defined by a 10\,\% decrease from the value above \Tc\ to the lowest-temperature value.  Optimal doping in this system corresponds to a measured Rh content of 0.09\footnote{Note that optimal doping is lower in \Eu\cite{Jiang2009,Jiao2017} than in the BaFe$_2$As$_2$ family\cite{Ni2009}.}, obtained for nominal concentrations of 0.12, so the ``nearly optimal'' samples investigated here are well within the uncertainty of optimal doping.  This paper uses the nominal concentrations.

Angle-resolved photoemission spectroscopy (ARPES) was utilized to measure the electronic structure of the samples. Besides traditional ARPES with a helium lamp ($h\nu$ = 21.2\,eV, resolution 10\,meV), a laser-based ARPES system ($h\nu$ = 6.3\,eV, resolution better than 5\,meV) with a micrometer-scale focal spot ($< 5 \mu$m)\cite{Iwasawa2017} was also utilized.  Since such a highly-focused beam can lead to space charge effects\cite{Zhou2005,Liu2008}, it was necessary to determine the maximum laser fluence that would avoid charging, then collect data below this intensity.  All ARPES measurements were performed at a base pressure better than $5 \times 10^{-11}$\,mbar.  ARPES was performed on a separate sample from the same batch.

\section{Results and Discussion}

The electronic structure of \EuRhopt\ measured by ARPES with a helium lamp ($h\nu$ = 21.2\,eV) is displayed in Fig.~\ref{fig:FSandBand}. Figs.~\ref{fig:FSandBand}(a-d) show the Fermi surface and constant-energy contours at binding energies $E_b$~=~0, 20, 40, and 60\,meV taken at a temperature of 15\,K, just below the cricital temperature. Similar to \Eu\ with other dopants\cite{Thirupathaiah2011,Xia2014}, there are hole pockets around the $\overline{\Gamma}$ point and electron pockets around the $\overline{M}$ points. The band dispersions centered around the $\overline{\Gamma}$ point [along cut \#1 in Fig.~\ref{fig:FSandBand}(a)] and $\overline{M}$ [along cut \#2 in Fig.~\ref{fig:FSandBand}(a)] at different temperatures are shown in Figs.~\ref{fig:FSandBand}(e-h) and Figs.~\ref{fig:FSandBand}(i-l), respectively. Energy distribution curves (EDCs) are shown as a function of temperature at Fermi momenta near $\overline{\Gamma}$ and $\overline{M}$ in Figs.~\ref{fig:FSandBand}(m) and \ref{fig:FSandBand}(n), respectively.  Upon increasing temperature through the critical temperature \Tc\ (from 15 to 50\,K), no significant changes in the band structure can be observed around the $\overline{\Gamma}$ or $\overline{M}$ point, we do not observe a gap opening, and no features near the Fermi level $E_F$ can be unambiguously identified as coherence peaks.  This is most likely a consequence of measuring too close to \Tc, where the gap and coherence peaks may not have fully developed.

\begin{figure}[htb]
  \includegraphics[width=0.8\columnwidth]{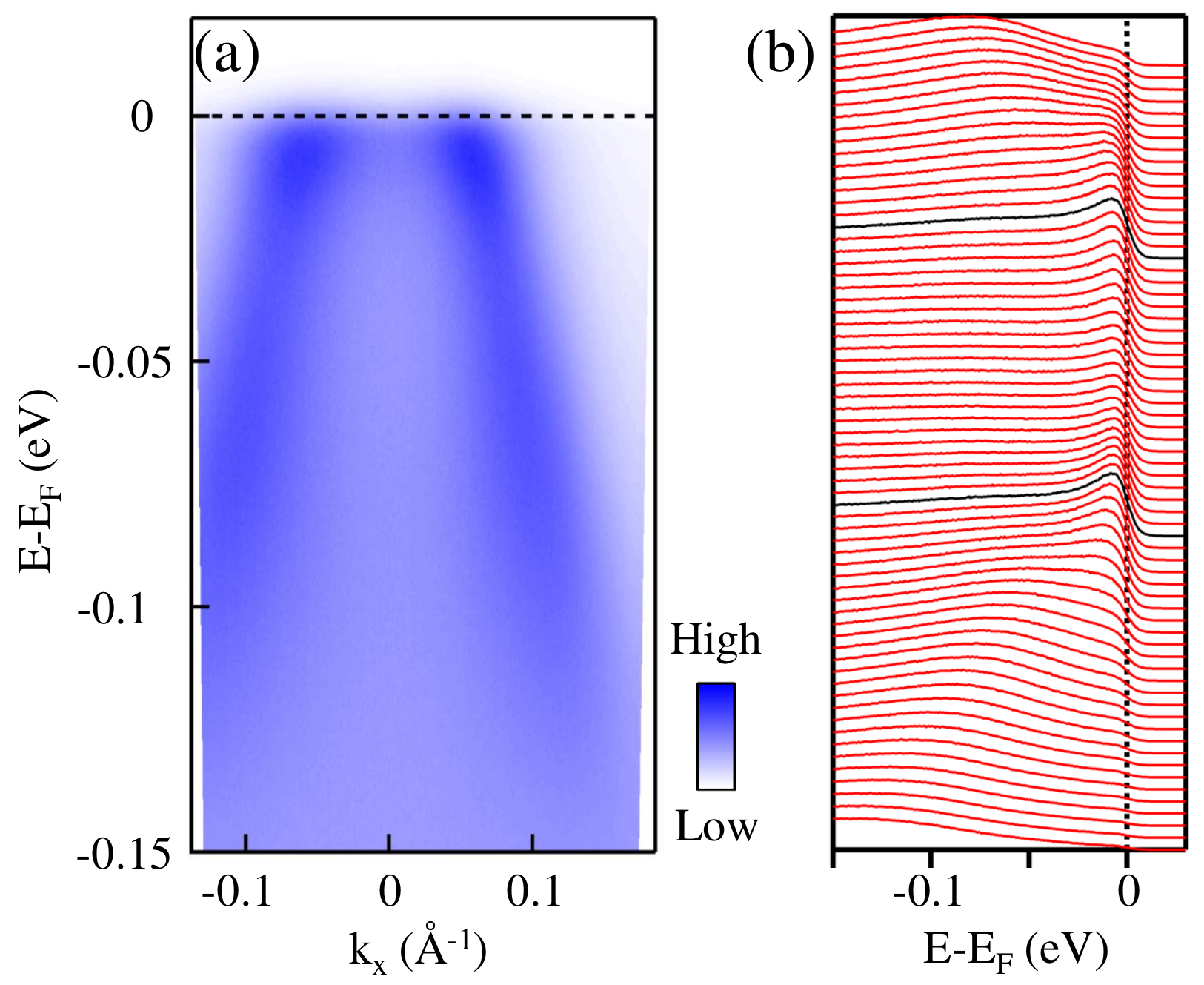}
  \caption{\label{fig:LaserBand} (a) Band structure of \EuRhopt\ crossing the $\overline{\Gamma}$ point along the $\overline{M}$--$\overline{\Gamma}$--$\overline{M}$ direction measured by $\mu$-laser ARPES ($h\nu$ = 6.3\,eV) at 13\,K. 
    (b) Energy distribution curves (EDCs) extracted from (a). EDCs at the Fermi momenta are plotted in black. 
  }
\end{figure}

To improve the energy resolution and avoid any inhomogeneity of the cleaved surface, laser-ARPES with a spot size below 5~$\mu$m was also used to measure the band structure around the $\overline{\Gamma}$ point (Fig.~\ref{fig:LaserBand} shows data at 13\,K). Despite an energy resolution better than 5\,meV, there are still no significant changes with temperature and no visible gap. 
A superconductor with a high transition temperature would normally have a sizeable gap, but this close to the transition temperature it would be very small.  The lack of observable signatures of a pairing gap may also arise from space charge effects, which we cannot completely rule out in the laser-ARPES data, or Eu acting as a magnetic impurity.  The pair-breaking induced by magnetic impurities can lead to ``gapless superconductivity''\cite{Abrikosov1960,Phillips1963}, a state which was discussed extensively in connection with BCS superconductors.

The band dispersion measured at 13\,K in the s polarization geometry (electric field perpendicular to the plane of incidence) is shown in Fig.~\ref{fig:LaserBand}(a). This geometry will selectively detect $d_{xy}$ and $d_{yz}$ orbitals, for polarization along $y$ with $z$ being the surface normal. The difference in the Fermi wavevector $k_F$ between Figs.~\ref{fig:LaserBand}(a) and \ref{fig:FSandBand}(e-h) is mostly likely due to either $k_z$ dispersion or differences in carrier doping at the surface --- the material cleaves through the Eu charge reservoir layer. As seen more clearly in 
the EDCs shown in Fig.~\ref{fig:LaserBand}(b), there is a feature suggestive of a flat band near the Fermi level, reminiscent of heavy fermion physics.  Such a feature can arise from a flat $4f$ band interacting with the regular electrons at low temperature (and having dipole symmetry that matches $d_{xy}$ and $d_{yz}$), or it can appear when a band top or bottom is just above the Fermi level.  A parabolic fit indicates that the top of the hole pocket at $\overline\Gamma$ should lie $\sim$30\,meV above the Fermi level, and this apparent flat band is most likely a tail from that band top, cut off by the Fermi function. 

\begin{figure}[htb]
  \includegraphics[width=0.75\columnwidth]{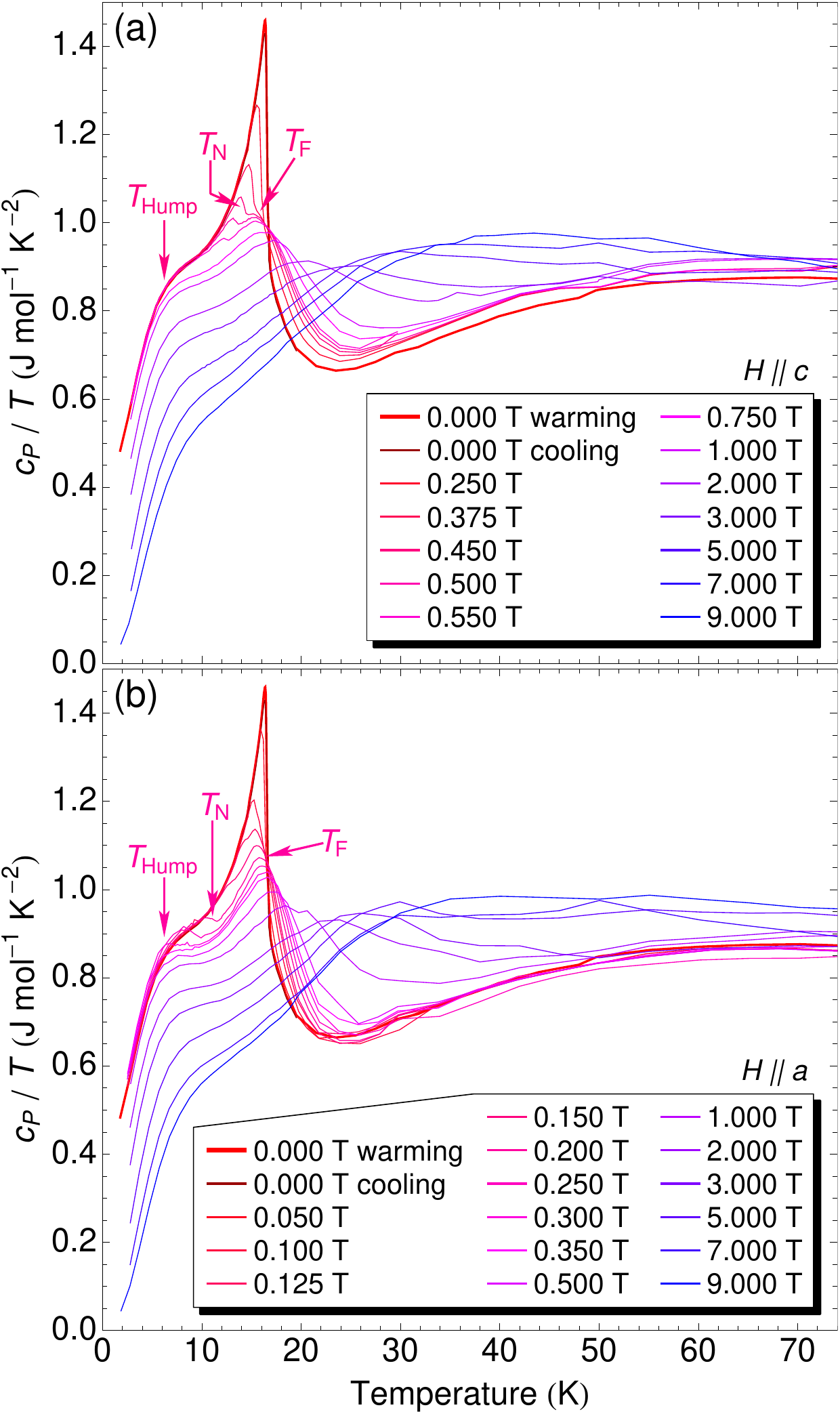}
  \caption{\label{fig:cP}Specific heat of \EuRhopt\ for fields (a) $H\parallel c$ and (b) $H\parallel a$.  The observed transitions are highlighted for the 0.45\,T trace in (a) and 0.20\,T in (b). \TN\ is the N\'eel transition and the hump at $T_F$ represents field-induced ferromagnetism.}
\end{figure}

\begin{figure*}[htb]
  \includegraphics[width=\textwidth]{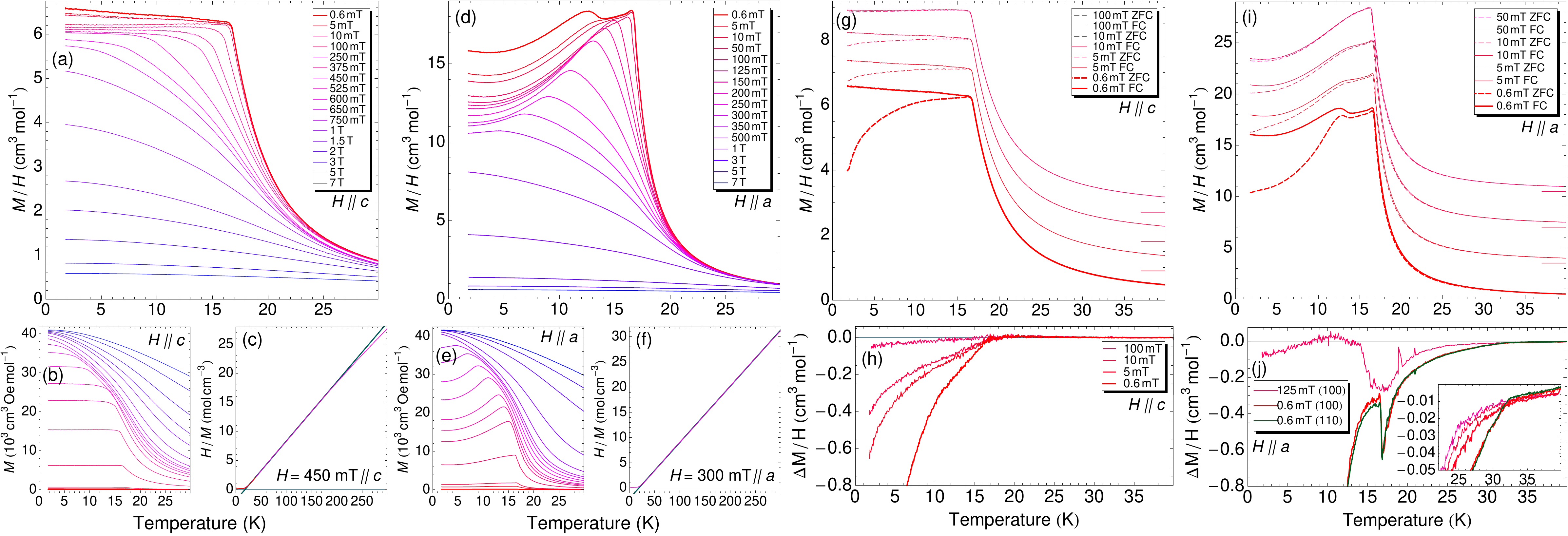}
  \caption{\label{fig:M}Field-cooled susceptibility $M/H$ of \EuRhopt\ for (a) fields along the $c$ axis, with the Curie-Weiss-like inverse susceptibility shown in panel (c).  The field-cooled susceptibility for an in-plane field is shown in (d), and its inverse in (f).  (b) and (e) replot the data in (a) and (c), respectively, as $M$ rather than $M/H$, to show the moment saturation.  (g) Comparison between the field-cooled and zero-field-cooled susceptibility for small $c$-axis fields, and (h) their difference.  (i) Field-cooled and zero-field-cooled susceptibility for selected in-plane fields and (j) their difference for two fields.  Data in panels (g) and (i) have been shifted vertically for clarity; the new zero positions are marked by horizontal lines.  The much larger susceptibility values for in-plane fields indicate moments oriented primarily along $c$.}
\end{figure*}

The specific heat $c_P$ of \EuRhopt\ is plotted as $c_P(T)/T$ vs.\ $T$ in Fig.~\ref{fig:cP}a for fields along the $c$ axis, and in Fig.~\ref{fig:cP}b for in-plane fields, and the susceptibility $M/H$ is plotted in Fig.~\ref{fig:M}.  In the zero-field specific heat, a sharp spike around 16.5\,K containing considerable entropy marks a phase transition, while a second large hump is visible around 6\,K.  Data taken on warming and cooling in zero field do not exhibit clear evidence of hysteresis.  The considerable entropy released below 20\,K would make \EuRhopt\ a heavy fermion superconductor if this were electronic entropy.  However, the superconducting transition in \EuRhopt\ nearly coincides with a magnetic ordering temperature, so a significant fraction of this entropy would be expected to be magnetic. Field dependence is invaluable for disentangling these origins.  The low-temperature hump gradually broadens and moves to higher temperatures with field, while the spike at the transition splits at low fields.  The bulk of the entropy from the 16.5\,K transition broadens and moves to higher temperatures, while a sharp jump moves rapidly to low temperature.  This behavior is qualitatively the same for both field orientations, but the rates at which the transitions move differ.

Since the vast majority of the entropy moves to higher temperatures in field, the magnetic contribution evidently dominates.  The sharp jump that moves rapidly to lower temperatures in field could in principle be either superconducting or antiferromagnetic. Susceptibility, shown in Fig.~\ref{fig:M} shows a clear transition matching the sharp jump in the specific heat, but because full diamagnetism is not achieved, it is unable to clarify whether this is magnetic or superconducting.  The rapid shift to lower temperature is inconsistent with the very gradual field dependence of the resistive superconducting transition\cite{Jiao2017}, suggesting an antiferromagnetic origin, but we note that resistive measurements may not be bulk-sensitive if superconductivity is present.  In nonsuperconducting Ca-doped \EuCaopt, an antiferromagnetic transition observed in the susceptibility at 15\,K is suppressed to zero by a 1.5\,T field applied along the $c$ axis, or by a 0.75\,T in-plane field \cite{Tran2018}.  The suppression of the specific heat jump in Fig.~\ref{fig:cP} has slightly higher anisotropy and is suppressed by somewhat lower fields, but its response to field is remarkably similar.  Based on the similarity to the Ca-doped material and the lack of any other sharp magnetic transition in a material which is known to magnetically order, we tentatively attribute the sharp jump to a bulk antiferromagnetic transition.

In magnetic field, the Ca-doped material\cite{Tran2018} and high-pressure-synthesized \Eu\cite{Tsvyashchenko2013} exhibit a transition from the paramagnetic normal state to field-induced ferromagnetic order, which moves rapidly to higher temperature as field is applied.  The broad specific heat hump that grows out of the sharp phase transition in \EuRhopt\ has a similar field dependence.  As seen in Fig.~\ref{fig:M}, our susceptibility data begin to deviate at progressively higher temperatures with field, and this can be clearly identified in the magnetization as a saturation.  This supports an interpretation of field-induced ferromagnetism for the higher-temperature specific heat hump in \EuRhopt.  

Our susceptibility measurements do not identify any feature that would correspond to the low-temperature hump in the specific heat.  The moments have minimal entropy below this temperature, but begin to develop some above.  This would indicate that the hump corresponds to stronger locking-in of moments or freezing out of their fluctuations at low temperature.  In optimally-Ir-doped \EuIropt, tilting of the in-plane Eu spins toward the $c$ axis sets in about 2.5\,K below the magnetic ordering temperature\cite{Jiao2013}, but the higher susceptibility values for in-plane fields in the Rh-doped material indicate moments that already lie primarily along $c$, thus a similar reorientation is unlikely here. 

Field-cooled (FC) and zero-field-cooled (ZFC) susceptibility data on \EuRh{0.12}{0.88}\ are presented in Figs.~\ref{fig:M}(g) and \ref{fig:M}(i) for low fields along the $c$ and $a$ axes, respectively.  In most doped \Eu\ materials, particularly in in-plane fields\cite{Matsubayashi2011,Jiao2013,Tsvyashchenko2013,Baumgartner2017,Tran2018}, the material does not exhibit full diamagnetism in the superconducting state, only a diamagnetic shift on top of a large paramagnetic background.  In general, a field-trained difference can be explained by either the freezing-in of magnetic moments or by superconductivity with vortex pinning.  The difference between FC and ZFC susceptibility data is commonly used in these materials as an indication of the superconducting contribution. 

A small difference between zero-field-cooled (ZFC) and field-cooled (FC) susceptibility is indeed visible in the lowest fields, with ZFC lower as expected; this difference is plotted in Figs.~\ref{fig:M}(h) and (j).  For $c$-axis fields, its onset at $\sim$17\,K is nearly field independent, but it becomes difficult to detect for fields above 100\,mT.  A similar difference persists roughly twice as high in temperature for in-plane fields, to approximately double the antiferromagnetic transition. A locking-in of spin moments in the paramagnetic state above the bulk magnetic transition is exceedingly unlikely, particularly since the Eu layer has not been directly disordered by cation substitution, so this may represent superconductivity. However, superconductivity to such a high temperature has never been observed in the \Eu\ system, including in resistivity measurements on similar samples\cite{Jiao2017}.  The onset temperature of $\sim$32\,K is approaching the maximum \Tc\ reported on doping the nonmagnetic analogs \Ba\ and \Sr, so stacking faults or Eu-deficient layers may be responsible.  Note that identifying this downturn with superconductivity would also imply a similar superconducting component in the undoped parent compound, where we observe similar behavior [see Figs.~\ref{fig:MAll}(f) and (h)], although this is nearly an order of magnitude weaker and is suppressed by much lower fields.  While in-plane fields may disrupt or reorient precursor magnetic order, and our thin platelet superconductor may also benefit from the thin limit in this orientation, we consider it unlikely that this feature is representative of the bulk of the sample. 

\begin{figure}[htb]
  \includegraphics[width=0.75\columnwidth]{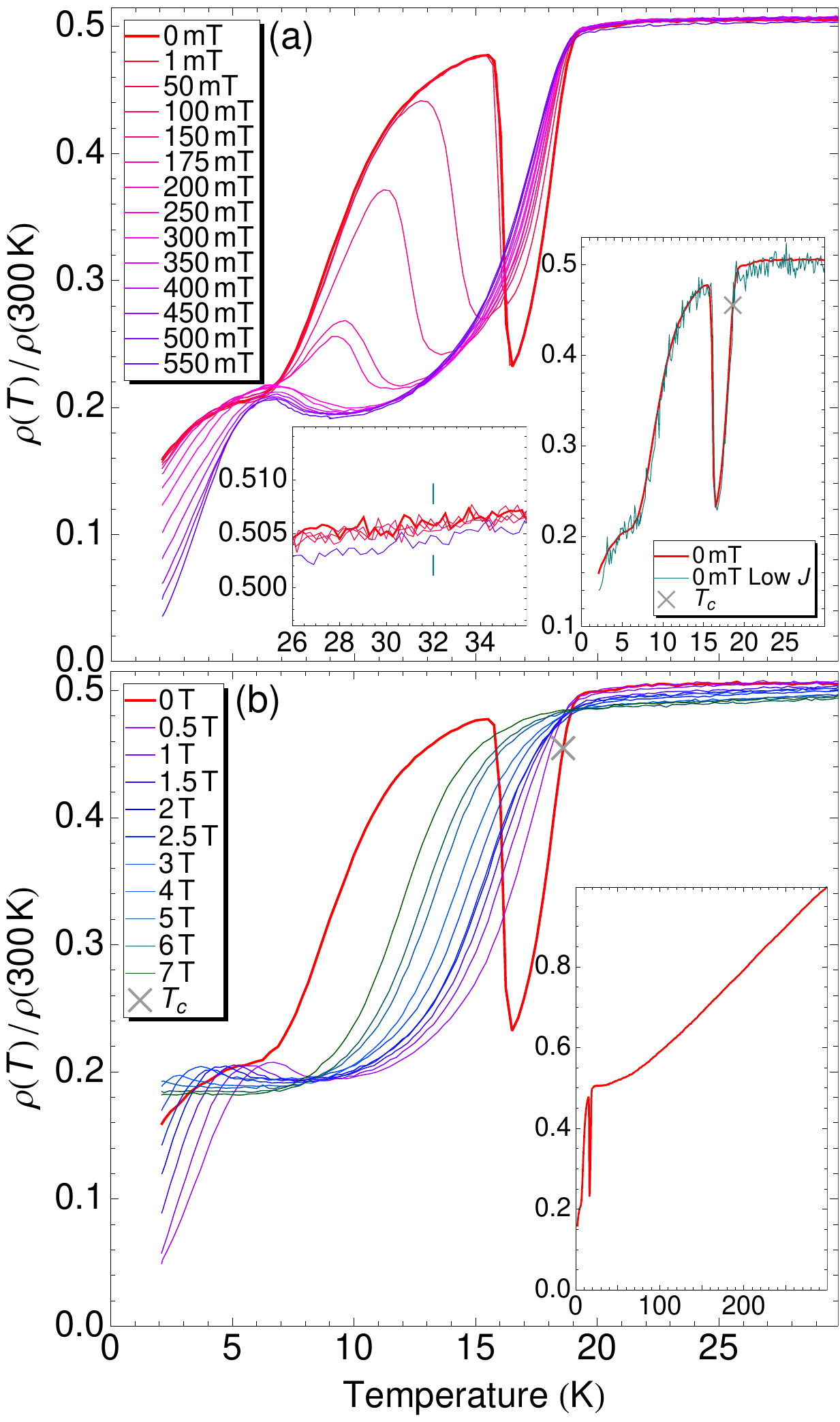}
  \caption{\label{fig:R}Resistivity of \EuRh{0.12}{0.88} for $H\parallel a$. (a) Resistivity for low fields, showing the suppression of the magnetic transition.  The left inset highlights the region around 32\,K where a feature was observed in the magnetic response for the lowest four fields and 0.55\,T, while the right inset shows the effect of reducing the drive current by a factor of 25.  (b) High-field data, showing the gradual suppression of superconductivity.  The inset shows the zero-field data up to 300\,K.  \Tc, defined as described in Section~\ref{expt} (Experimental) above, is marked with a gray cross in (b) and the larger inset to (a).}
\end{figure}

To check whether the apparent $\sim$32\,K onset could be superconductivity, we measured the resistivity of a \EuRh{0.12}{0.88} crystal from the same batch in in-plane fields, as shown in Fig.~\ref{fig:R}.  Magnetization and EDX data [$x$=0.097(6)] on this crystal are indistinguishable from those presented above.  As seen in the left inset in panel (a), there is no clear feature around 32\,K; the first hint of a downturn is around 22\,K.  As shown above, small fields suppress the magnetic transition, and this visibly reduces the low-temperature resistivity.  Increasing the field beyond 0.5\,T then gradually suppresses the main downturn from $\sim$19\,K to $\sim$15\,K at 7\,T.  Its gradual suppression by field and obvious competition with the magnetic order allow us to identify it as superconductivity. A second transition around 7\,K which is suppressed below 1.8\,K by a 6\,T field is most likely due to vortex pinning, since it is not observed by any other technique.  The resistivity does not approach zero, so the initial downturn is taken as indicative of the superconducting transition.  Aligning the field exactly perpendicular to the field is very easy in the susceptibility measurement, but more difficult in the geometry used for our resistivity measurements, so our inability to observe any hint of a downturn at higher temperatures could be a result of inexact field alignment.  However, if a misalignment of perhaps a few degrees were able reduce the onset temperature by half, that would be difficult to understand if it were bulk superconductivity. A test in which the field was rotated within the $ab$-plane found no angle dependence at low temperature, suggesting that perfect in-plane field alignment is not crucial.

We note at this point that the physical properties of these crystals differ from those with identical nominal dopings in Ref.~\onlinecite{Jiao2017}, most noticably the zero-field resistivity below \Tc. The behavior below \Tc\ in \Eu-based materials is determined by an intricate interplay between superconductity and magnetism, resulting in reentrant resistivity in most systems, which is explained in terms of flux-flow resistivity in a spontaneous vortex state\cite{Ren2009,Jiao2011,Jiao2017,Jiao2017b}.  In this scenario, the exact shape of the resistivity reentrance is highly sensitive to the sample quality, including defects, dopant homogeneity, and pinning sites, and even to the direction along which the current is applied.

Similar to the reentrant resistivity, the difference between ZFC and FC susceptibility also shows a sharp jump back toward zero for in-plane fields, coinciding with the peak in the susceptibility.  This is a result of the sample temperature lagging the thermometry --- tests varying the temperature scan rate and comparing zero-field-cooled warming with field-cooled cooling and field-cooled warming measurements are shown in Fig.~\ref{fig:test} in the Supplemental Materials\cite{SMref}.  This jump in the difference between ZFC and FC data is only seen for in-plane fields because only in-plane fields have a nonzero difference between ZFC and FC data at this temperature.  

The inverse susceptibility, shown in Figs.~\ref{fig:M}c and \ref{fig:M}f, is well fit by the Curie-Weiss law [$M/H = C/(T-\Theta_W)$] over a wide temperature range, and indicates predominantly ferromagnetic interactions with minimal frustration.  The extracted Weiss temperatures are 18.5 and 19.4\,K for $c$-axis and in-plane fields, respectively, consistent with that reported earlier for the parent compound\cite{Raffius1993}.  These are only marginally higher than the bulk transition in the susceptibility and specific heat, indicating minimal frustration.  The respective extracted moments of 8.85 and 8.46\,$\mu_B$ for $c$-axis and in-plane fields exceed that expected for spin-7/2 Eu$^{2+}$ (7.95\,$\mu_B$), presumably due to a contribution from Fe spins.  However, the shape in $M(T)$ and field-suppression of the transition are consistent with antiferromagnetic order, and this remains true at the higher Rh doping of 0.16\cite{SMref}.  In Co-doped \Eu, the magnetic ground state of the Eu ions evolves from an A-type antiferromagnet, to a canted antiferromagnet, and eventually to a ferromagnet\cite{Jin2016} with doping, and a similar transition to ferromagnetism is observed with P or Ir doping\cite{Nowik2011,Anand2015}. Our results indicate that we do not reach a high enough doping to obtain a ferromagnetic ground state.

Significantly larger fields are required to suppress the antiferromagnetic transition when the field is applied along $c$ than $a$.  Since it has been previously reported that in the undoped parent compound and at low dopings for various dopants the spins lie in the plane and order ferromagnetically within each Eu plane\cite{Xiao2009,Nowik2011,Jin2016}, this indicates that it is relatively easy to realign the spin orientation of the Eu slabs, thereby breaking the global antiferromagnetic order.

A hump at 12-14\,K in the in-plane susceptibility at low fields in P- and Ir-doped material\cite{Zapf2013,Baumgartner2017} has been attributed to a spin-glass transition below the antiferromagnetic transition in analogy to EuFe$_2$P$_2$. The picture here is that the $c$-axis-aligned ferromagnetic moments have a disordered $ab$-plane component driven by antiferromagnetic interlayer RKKY interactions, and this in-plane component freezes. The freezing of a perpendicular spin component has also been discussed in Ref.~\onlinecite{Jiao2017} for Rh-doped material. The remarkable agreement of our susceptibility with the Curie-Weiss law down nearly to the transition and the lack of any other sign of frustration suggest that this component is very small if it exists, but the present data do not allow us to exclude it as a possibility.  This hump shifts to higher temperatures with field, but its strength in $M$ is roughly constant, making it vanish rapidly in $M/H$ as field is increased.  Its strength also apparently grows with Rh doping\cite{SMref}.  This behavior would be most consistent with a saturated magnetic moment associated with the dopant site.  Its consistent appearance in multiple single-crystal studies by different groups using different FeAs-layer dopants supports this interpretation, but demonstrating similar doping dependence in the other systems would provide important confirmation.

As part of this work, we also measured the susceptibility of the parent compound and several Rh dopings, but found minimal differences among the doped samples\cite{SMref}.  Evidently, adding enough Rh dopants to induce superconductivity significantly reduces the fields required to suppress the Eu magnetic order, and slightly reduces its onset temperature.  The Eu layers are coupled through the (Fe,Rh)As layers, which also order in the undoped parent compound.  Rhodium doping suppresses the Fe-site magnetic order, which may directly weaken the Eu interlayer coupling.  The singlet superconducting ground state in the FeAs planes should also be less efficient at RKKY coupling.  The proximity to strong rare-earth magnetism will ensure that some carriers remain unpaired, so some weak interlayer coupling remains, and the Eu moments continue to exhibit long-range order, but the reduced interlayer coupling makes it more fragile.  Evidently, once there is enough Rh present to disrupt the Fe magnetism and reach a superconducting ground state, tuning this ground state has no further effect on the Eu moments.

\begin{figure}[htb]
  \includegraphics[width=0.9\columnwidth]{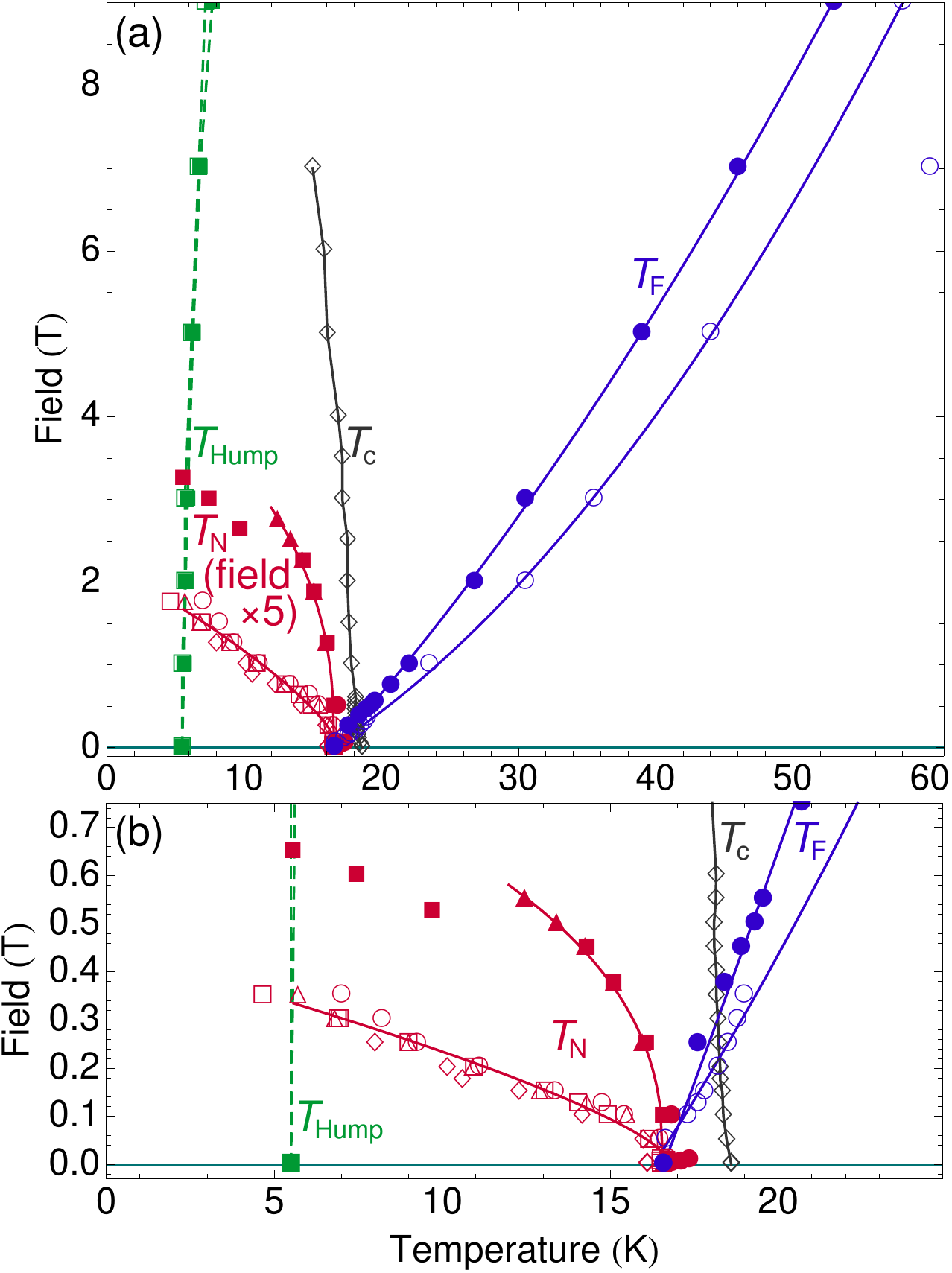}
  \caption{\label{fig:HT}(a) $H$--$T$ phase diagram of \EuRhopt\ obtained from our specific heat and susceptibility data. Filled symbols are for $H\parallel c$, and open symbols are for $H\perp c$; lines are guides to the eye. $T_F$ represents field-induced ferromagnetism, $T_\text{Hump}$ identifies the low-temperature hump in the specific heat, $T_c$ marks the resistive superconducting transition, and $T_N$ represents the bulk antiferromagnetic transition.  $T_N$, the field of which has been expanded by a factor of 5 for clarity, includes points based on the specific heat jump (triangles), susceptibility peak (squares), resistivity (diamonds), and susceptibility jump (circles). (b) Expanded view of the low-field region around \TN\ and \Tc; here \TN\ has not been expanded.}
\end{figure}

Based on the specific heat, resistivity, and susceptibility data, it is possible to assemble an $H$--$T$ phase diagram for \EuRhopt, shown in Fig.~\ref{fig:HT}.  The position of the lower-temperature hump in the specific heat was estimated based on its peak position, the antiferromagnetic transition was determined based on the midpoint of the specific heat jump, and for the high-temperature specific heat hump a midpoint was estimated.  The peak position and midpoint of the jump in the susceptibility were added, and resistive transitions are based on a 10\,\% decrease from the value above \Tc\ to the lowest-temperature value and the midpoint of the jump at \TN. The phase diagram (Fig.~\ref{fig:HT}) combines the results of our measurements on \EuRh{0.12}{0.88} for both orientations. The antiferromagnetic transition has been expanded by a factor of 5.  Aside from the inclusion of superconductivity, this phase diagram bears a very close resemblance to that of nonsuperconducting Ca-doped \EuCaopt\cite{Tran2018}.  Such a high degree of similarity is surprising given that the dopants in the latter case are on the Eu site, disrupting the magnetic order, while here the Rh is in the superconducting FeAs planes, weakening the competing superconductivity. 

Since all features in the specific heat near the superconducting transition can be attributed to magnetism, this raises the question of why no trace of the superconductivity is visible.  The specific heat jump at \Tc, $\Delta c_P/\Tc$, in the nonmagnetic BaFe$_2$As$_2$ family scales with \Tc\cite{Budko2009}, and values in the SrFe$_2$As$_2$ family are quantitatively similar\cite{LeitheJasper2008}.  If the FeAs layers are assumed to behave the same in the Eu material, similar scaling should apply, which would predict a jump height of roughly 10\,mJ/mol$\cdot$K$^2$, less than 2\,\%\ of the specific heat near the transition.  The uncertainty on the specific heat data near \Tc\ is less than 1\,\%, but at low fields the data at that temperature are increasing precipitously toward a phase transition.  It would be difficult to detect such a small jump within that underlying background. Any broadening of the transition, which would be expected given the disorder and inhomogeneity that must accompany the replacement of Fe atoms with Rh, would make it even more challenging.  At fields on the order of 0.5\,T, the magnetic specific heat is far less singular, but there is still no superconducting transition visible.  This is a small change in field for the superconductivity, so the superconducting transition should only be very slightly weaker and broader. In fact, the lower resistivity around this field suggests that we might even expect a {\slshape stronger, sharper} superconducting transition with the magnetism suppressed. It should be visible, and it is not.

The key may lie in the competition with strong rare earth magnetism.  The magnetic fields impinging upon the FeAs layer even in zero external field will induce a vortices and shielding currents\cite{Jiao2017b}, reducing the entropy saved in the superconducting state.  The specific heat of a Type-II superconductor in field has a reduced jump height, which may push it beyond our detection limit even in low or zero field in doped \Eu. We note that no specific heat jump at \Tc\ has been observed in Ir-doped and P-doped Eu-122 either\cite{Paramanik2014,Nandi2014}, likely for the same reasons.  Unfortunately, this makes it difficult to confirm that the observed superconductivity is an intrinsic, bulk property.  If the material is effectively in high field even at zero applied field, that could offer another alternative explanation for the absence of a pairing gap in ARPES.  Under high (effective) fields in a Type-II superconductor, the gap is not only reduced but also spatially inhomogeneous on a lengthscale well below that of even our $\mu$-ARPES spot size.  This would broaden and fill in the gap, making it impossible to observe.  

Since the superconductivity is forced to coexist with strong rare-earth magnetic moments even in their paramagnetic state, the superconducting phase exists in an intrinsic vortex state even at low fields\cite{Jiao2017b}.  This will reduce not only the entropy savings, but also the diamagnetic shielding, making the superconductivity largely invisible to both specific heat and magnetization.  Indeed, no unambiguous signatures of superconductivity are immediately obvious in our data for any doping or field orientation.

We now return to the problem of the susceptibility glitch at 32-34\,K in in-plane fields, suggestive of superconductivity. This was not observed in the resistivity, for which several explanations are possible.  One possibility would be that the material is in a regime of free flux flow, so that the drive current for the resistivity measurement moves the vortices perpendicular to the current, causing resistive losses. Reducing the drive current by a factor of 25 [Fig.~\ref{fig:R}(a) inset] had no significant effect, but this could still be too large for very weak pinning. Another explanation could be that this higher-temperature superconducting contribution may be destroyed by a small misorientation of the field.  A 45$^\circ$ rotation of the field within the $ab$ plane had no effect on the data (not shown), implying that any such misorientation would need to introduce a $c$-axis component; however, a rotation of the field about the $c$-axis, which should produce perfect in-plane alignment at some angle, found no angle-dependence.  Superconductivity ocurring only in low in-plane fields could suggest that disrupting the antiferromagnetic Eu order greatly enhances superconductivity, but it could also be consistent with thin superconducting slabs, perhaps at the surface or at stacking faults, which are more readily observed under in-plane fields thanks to thin-limit effects.  The weak nature of this downturn and its appearance at the same temperature irrespective of doping, but only for one field orientation, suggest an extrinsic origin.  Nonetheless, the data are suggestive of superconductivity at roughly double the highest known \Tc\ in doped \Eu, and this clearly warrants further investigation.

\section{Conclusion}

The antagonistic relationship between the Eu and FeAs layers in \Eu\ disrupts both the long-range magnetic order and bulk superconductivity.  The consequence for the magnetism is that the ground state can be tuned by doping the FeAs layers, likely disrupting the interlayer coupling, and our work indicates that the magnetism's $H$--$T$ phase diagram can be traversed with laboratory magnets.  The superconductivity may also be tunable --- our data indicate that the FeAs layers feel a strong field even under zero applied field.  The upper critical fields in the iron-based superconductors are normally very high and difficult to access, but the \Eu\ family may make the high-field region far more accessible.  In particular, ARPES is a strictly zero-field technique due to the immediate destruction of momentum information by any applied magnetic field, so field-dependent ARPES is ordinarily completely impossible.  Our results suggest that in the \Eu\ system, ARPES can be performed at a higher effective magnetic field.  Tuning the magnetic layers, for instance by Ca or Sr doping, should return the system to lower effective field, allowing an effective-magnetic-field-dependent ARPES study and directly accessing the electronic structure under conditions that are ordinarily inaccessible to ARPES.

\begin{acknowledgments}
The authors thank M.L.\ Zhu, X.M.\ Wang, and the group of J.\ Zhao at Fudan University for measurement assistance. This work was supported by the National Natural Science Foundation of China (Grant No.~11674367) and the Zhejiang Provincial Natural Science Foundation (Grant No.~LZ18A040002).  DCP is supported by the Chinese Academy of Sciences through 2018PM0036.  The ARPES beamtime at the Hiroshima Synchrotron Radiation Center was granted under Proposal 17AG004.
\end{acknowledgments}

\bibliography{EuFe2As2}
\appendix
\renewcommand{\thefigure}{S\arabic{figure}}
\renewcommand{\thesection}{S\arabic{section}}



\section*{Supplemental Material:  Additional Susceptibility Results}

\begin{figure}[b!]
  \includegraphics[width=0.75\columnwidth]{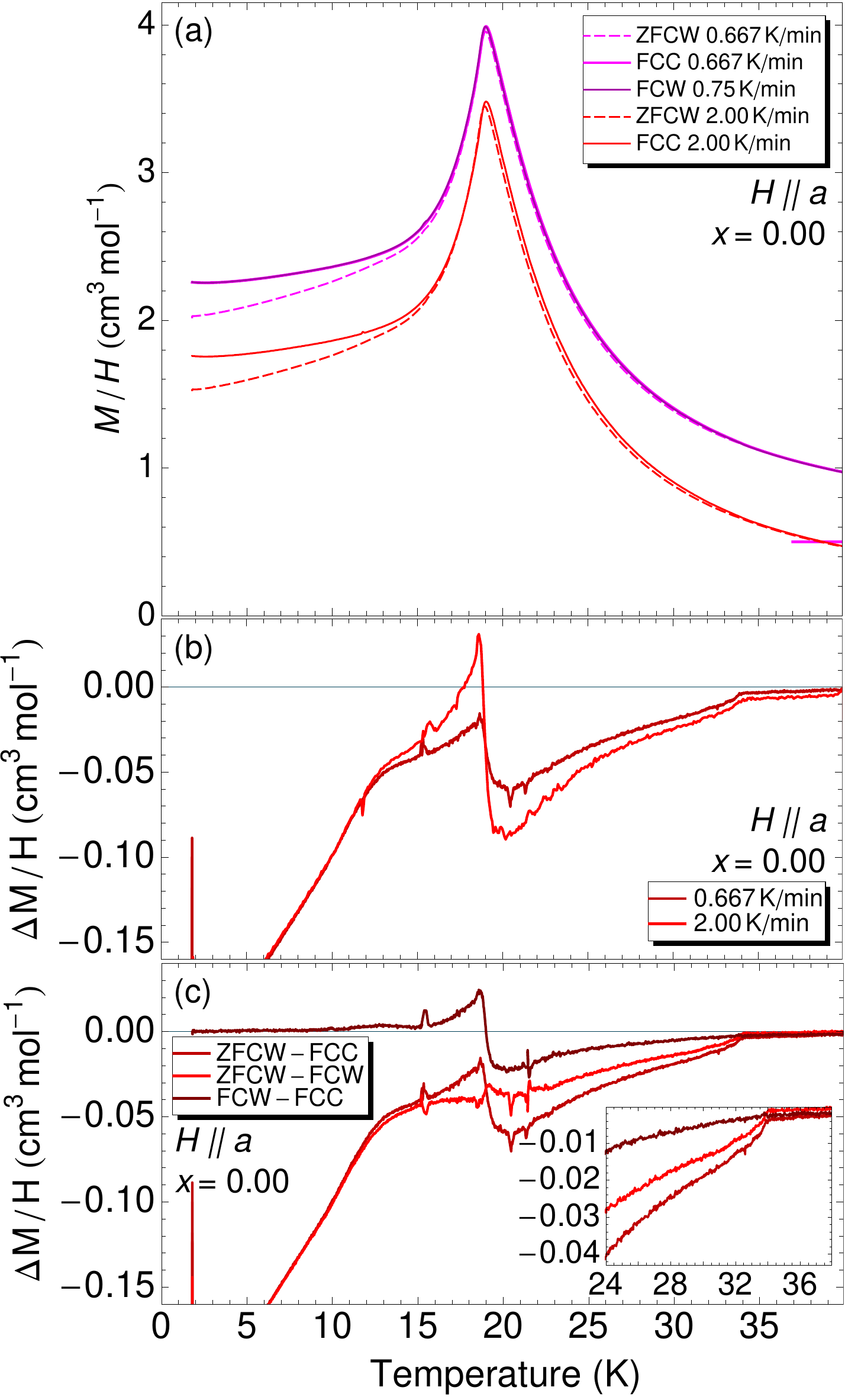}
  \caption{\label{fig:test}Effect of ramp rate and sweep direction on difference between field-cooled and zero-field-cooled data in undoped \Eu\ in an in-plane field of 0.7\,mT. (a) Susceptibility data.  Zero-field-cooled warming (ZFCW), field-cooled cooling (FCC) and field-cooled warming (FCW) are shown and sweep rates are labelled.  (b) Effect of sweep rate on difference between ZFCW and FCC.  (c) Comparing data taken at slower sweep rates, the glitch at the magnetic transition is absent if both ZFC and FC data are taken on warming, but the glitch around 34\,K is only present in ZFC data.}
\end{figure}

Figure \ref{fig:test} shows the results of the sweep rate and direction on the difference between zero-field-cooled and field-cooled susceptibility data.  As seen in Fig.~\ref{fig:test}(b), a slower sweep rate results in a smaller spike at the bulk transition, suggesting that this feature is an artifact due to the sample temperature lagging the thermometry.  Comparing field-cooled sweep directions in Fig.~\ref{fig:test}(c), this spike exists when cooling and warming are subtracted, but not when zero-field-cooled and field-cooled warming scans are subtracted, again confirming its extrinsic origin.  The glitch at 34\,K, however, exists independent of sweep rate or direction, and only when ZFC and FC runs are subtracted.  This indicates that it is a real effect, although not necessarily an intrinsic bulk property of the material.

\begin{figure*}
  \includegraphics[width=0.94\textwidth]{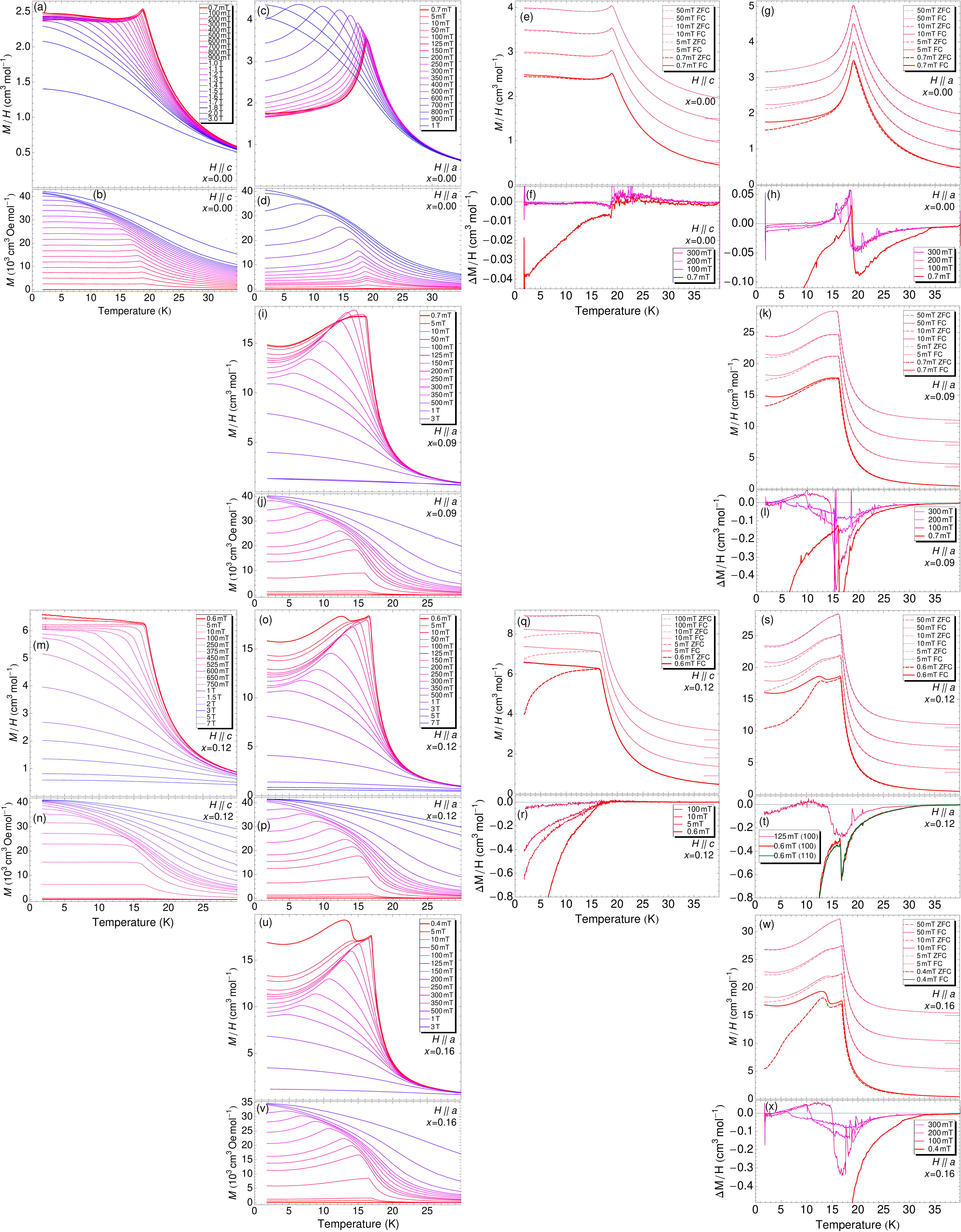}
  \caption{\label{fig:MAll}Field-cooled susceptibility $M/H$ of the undoped parent compound \Eu, for fields (a) along $c$ and (c) along $a$.  The corresponding magnetization data are shown in (b) and (d), respectively.  Zero-field-cooled and field-cooled susceptibility are compared for selected fields in (e) and (g), with their difference shown in (f) and (h), for fields along $c$ and $a$, respectively.  Equivalent data are presented for (i-l) \EuRh{0.09}{0.91}, (m-t) \EuRh{0.12}{0.88} and (u-x) \EuRh{0.16}{0.84}.}
\end{figure*}

It is useful to check the effect of Rh doping on the transition.  The physics in \EuRhx\ is dominated by a competition between strong rare-earth magnetism in the Eu layer and high-temperature superconductivity in the FeAs layer.  Doping with Rh is expected to have a strong effect on the superconductivity, while any impact on the magnetic layer will be mainly through changes in the interlayer coupling.  Magnetization measurements as a function of doping are presented in Fig.~\ref{fig:MAll}.  Adding Rh to the undoped parent compound changes the behavior somewhat, but further Rh doping only leads to nearly-imperceptible further changes.  The zero-field transition appears to shift to slightly higher temperature with doping, with a corresponding slight increase in the field required to suppress it, and the spin-glass hump around 13\,K appears to grow with doping.  In the parent compound, while a sharper peak is seen at the transition which looks similar and appears at slightly higher temperature than in the doped samples. This survives to significantly higher fields in the parent compound.  The parent compound is not believed to superconduct at ambient pressure.  This doping dependence provides additional evidence to suggest that the main transition here is antiferromagnetic ordering of the Eu moments, not superconductivity.


\end{document}